# Medical Applications - Instrumentation & Diagnostics

*Andreas Peters*
Heidelberg Ion-Beam Therapy Center (HIT), Department of Radiation Oncology, Heidelberg University Hospital, Heidelberg, Germany

**Abstract**
This CAS talk describes the role of beam instrumentation and diagnostics in particle therapy accelerators. It presents an extended view on instrumentation, feedbacks, detector technology, quality assurance (QA) and their interdependencies. Furthermore, some basics, examples and challenges in near future concerning diagnostics and instrumentation techniques used in particle therapy are reported.

## 1 Beam Instrumentation and Diagnostics in Particle Therapy Accelerators

There is no principal difference between a scientifically used accelerator and a medical accelerator. But due to the economic effects in the commercial field of particle therapy in hospitals sometimes only the absolute minimum of beam diagnostics is installed for commissioning and standard operation [1]. Most important devices are current transformers for online diagnostics (non-destructive) and beam profile monitors for the daily machine QA (destructive) – these topics are covered by other articles of these proceedings.

The specialty of medical accelerators is the strong link to the beam consumer, in this case the irradiation technology and the high-conformal dose application. These leads to an extended view of the topic, where instrumentation, feedbacks, detector technology and quality assurance of the medically used beam move closer together.

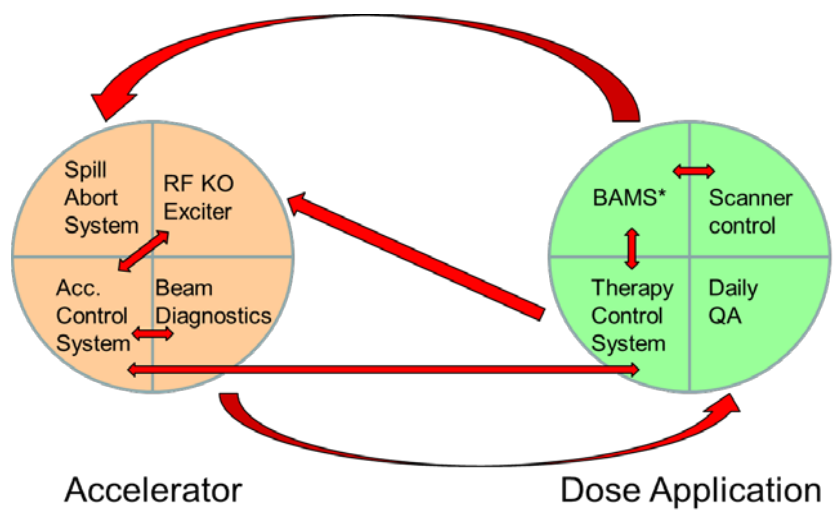

**Fig. 1:** Interactions of the accelerator part and the dose application in a particle therapy facility
(* BAMS: <u>B</u>eam <u>A</u>pplication and <u>M</u>onitoring <u>S</u>ystem)

Figure 1 shows exemplarily some of the tight interdependencies within and between the both 'worlds' of the accelerator and the dose application in a therapy facility. E.g. beam diagnostics in the accelerator is part of the daily QA of the dose application and uses the same techniques as the BAMS detectors, which are specially equipped ionization chambers (ICs) and multi wire proportional chambers (MWPCs), see chapter 2. As another example the accelerator instrumentation, here the RF knock-out exciter, is controlled by both worlds dependent on the execution mode of the facility, see chapter 3 for details. Thus, beam diagnostics and instrumentation of the accelerator form an integral part of the functionality of the whole beam application chain, see chapters 4 and 5 for further examples.

## 2     Interaction of Accelerator and Irradiation Technology in Particle Therapy

Raster scanning technique, also named pencil-beam scanning, is the actual method for dose application in proton and ion beam therapy. "Scanning" in the context of particle therapy is the application of the desired dose by means of a pencil beam, which is moved over the target point-by-point to cover the whole target volume [2]. This is in contrast to broad-beam-techniques that use an expanded particle beam to irradiate the whole area or volume of the target at the same time.

### 2.1     Basics of 3D-conformal Raster Scanning Technique

The part of a treatment plan corresponding to a fixed (water equivalent) depth is called an iso-energy-slice (IES). The selection of an IES is achieved by varying the beam energy. In case of a proton beam from cyclotrons with fixed end energy this is done using a degrader with changing efficiency due to the necessary deceleration. If the proton or ion beams are produced by a synchrotron, the energy variation can be actively done by stopping the accelerating ramp at the desired flattop energy.

Due to the shape of the Bragg peak, see Figure 1 (left diagram), not only the region where the particle stops is irradiated, but also the region before the peak ("plateau region"). This has to be taken into account during treatment planning. The dose plateau, result of overlapping beams with different energies, is called a spread-out Bragg peak (SOBP), see Figure 2 (right diagram).

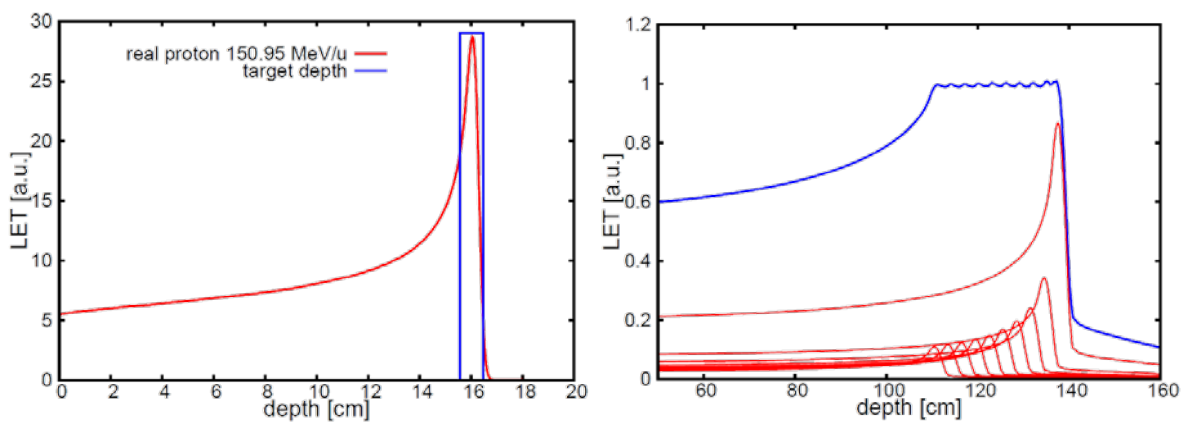

**Fig. 2:** Left diagram: The LET dose of a fixed energy proton beam deposited in water depending on the penetration depth showing the Bragg peak at the end; Right diagram: Overlapping of different Bragg peaks leads to a plateau, the so-called SOBP.

In a fully active system scanning in x-y-direction (transversal to the beam) is achieved by deflecting the ion beam in a pair of scanning magnets (horizontal, vertical). Typically, the beam will be kept at the designated position for each raster point until the desired dose for the point has been reached. Thus dose controlled irradiation is realized by using a feedback loop, which utilizes an ionisation chamber for measuring the beam current, see Figure 3 – an additional IC is needed due to redundancy for safety reasons. The dose is then deduced by numerical integration. In the same way two MWPCs determine the beam position and pass this information to the local control system, which can steer back the beam to its correct location in case of deviations.

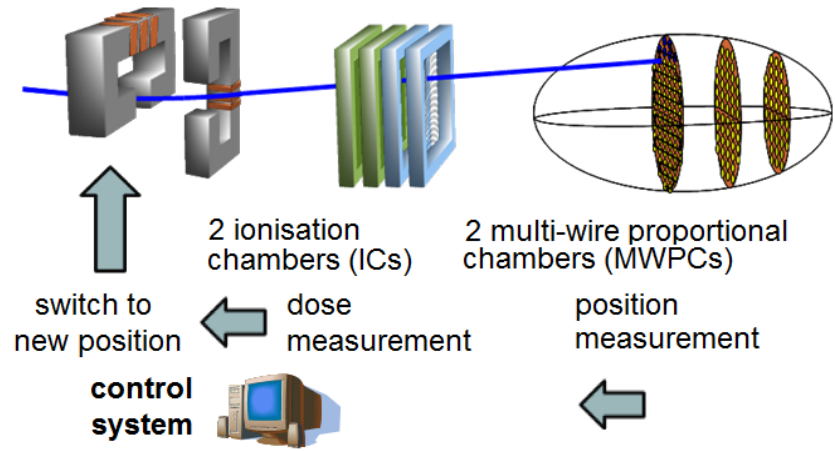

**Fig. 3:** Schematic view of fully active two-dimensional scanning system ionisation chambers for dose measurement and additional multi-wire proportional chambers for position control

In order to cover the whole cross-section of the tumour, a scan path (or "work list" of points) must be defined which determines the order in which different scan spots are irradiated. The spacing of the raster points has an enormous influence on the dose homogeneity: For a Gaussian beam shape a dose profile will become perfectly flat if the spot separation is equal or smaller than one sigma, see Figure 4. Dose profiles become clinically acceptable at a spacing of ~2σ. For treatment planning at HIT a spacing of 0.8σ or less is used as a rule of thumb to allow for some error in spot size.

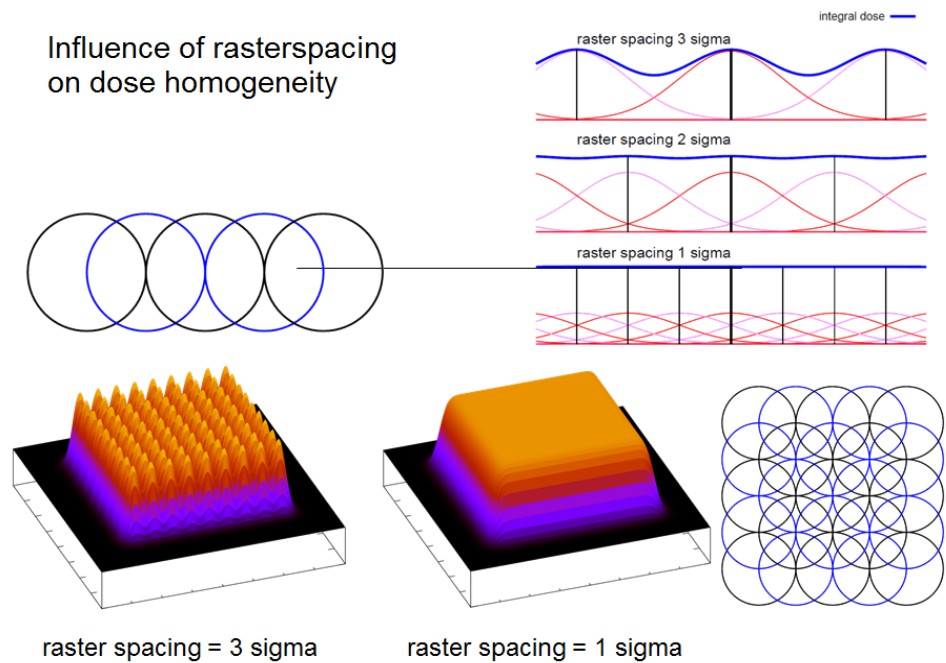

**Fig. 4:** The influence of the spacing of the raster points on the dose homogeneity

For a 3D-conformal dose application a further effect has to be taken into account: Putting together all IESs it is obvious that in the central region most of the dose is already applied by the distal IESs, at the lateral edges nearly the full dose is still missing see Figure 5. The variation in particle number in one single IES can be quite large, sometimes by a factor of ~100!

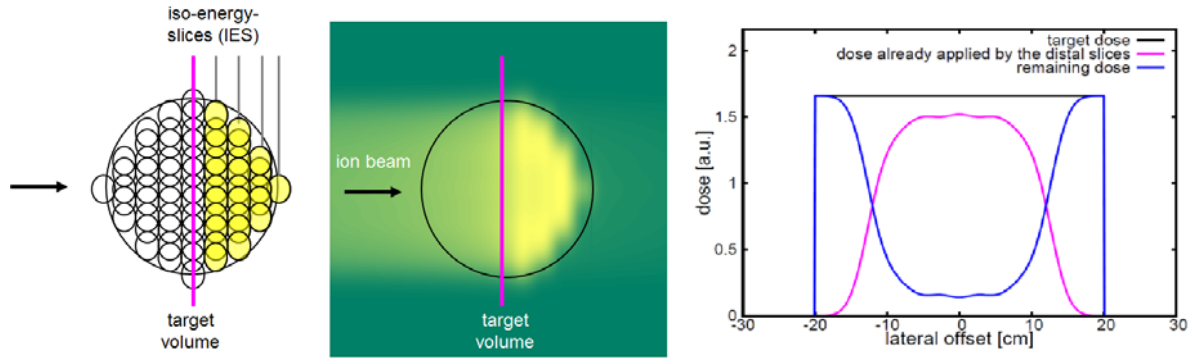

**Fig. 5:** The influence of the spacing of the raster points on the dose homogeneity

### 2.2 Implementation of the Raster Scanning Technique - Example: HIT

The raster scanning systems used in particle therapy nowadays have some special demands on the accelerator chain providing the necessary beams:

- Allowance to request different energies, beam spot sizes ("focus levels") and intensities from a pre-tuned and reliable library;
- Delivery of different combinations of all of these beam parameters within a few seconds;
- Delivery all of these combinations with a high beam quality sufficient for medical use;
- Provision of a spill pause functionality that means a possibility to interrupt the beam extraction in case of irradiation of disjoined raster point "isles" in one IES.

From these requirements and the demanded physical properties of the beams a list of parameters was deduced, which can be found in Table 1 [3].

**Table 1:** Beam parameters of the Heidelberg Ion Beam Therapy Centre (HIT)

| Parameter | Values and Remarks |
|---|---|
| Ions | protons and carbon, (oxygen) – 2 ion sources<br>In addition: helium (3. ion source; permission for patient treatment in 2020 awaited) |
| Intensity | $2 \times 10^6$/s to $8 \times 10^7$/s for carbon<br>$8 \times 10^7$/s to $4 \times 10^8$/s for protons<br>10 steps; maximum extraction time 5 s; spill pause possible |
| Energy | 88-430 MeV/u for carbon<br>50-221 MeV/u for protons<br>255 steps, 1-1.5 mm spacing, 2-30 cm range in water |
| Focus | 3.5-13 mm FWHM for carbon<br>11-33 mm FWHM for protons<br>4 steps |

Thus a total of 4 x 10 x 255 x 4 = 40,800 settings (max.) per treatment room have to be realized in the accelerator control system and filled by manual and semi-automatic beam tuning. In addition, up to 36 angles (10° steps) have to be handled at the gantry.

On the other side, the therapy control system adapting the dose to the individual patient treatment plan has to monitor the system on a timescale that matches the typical irradiation time of a raster point (~1-100 ms); this implies the following demands:

- Dose controlled irradiation system: dose measurement on a time scale that is short compared to the irradiation time of one point → here: 10 µs – faster than the drift time of ions in the ionization chamber;

- Position measurement with MWPCs on a time scale that is "short" compared to typical position variations (HIT: 250 µs);

- Beam position feedback system (linear) – see following text;

- Online monitoring of beam focus size and limit checks – a focus feedback loop is planned in the next generation control system, but for the implementation the (possibly changing!) beam optics have to be considered.

All these requirements are realized at HIT in the so-called BAMS, the Beam Application and Monitoring System see Figure 6. It consists of two (redundant) MWPCs for profile monitoring and position feedback and three (diverse and redundant) ionization chambers for dose measurement.

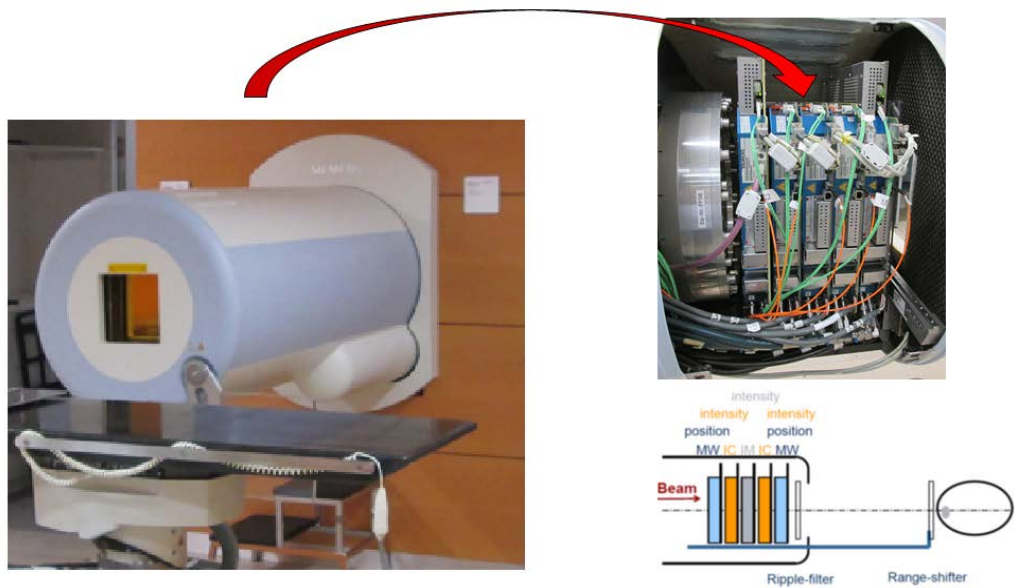

**Fig. 6:** The HIT-BAMS implemented in the nozzle (left) in the treatment rooms; the stack of detectors with attached data acquisition electronics (right).

The BAMS is an integral part of the raster scanning system. Further components are the scanner magnets (horizontal, vertical), where the beam is steered to hit the "right" voxel position in the iso-center plane (the tumour position in case of a patient treatment) – and a corresponding position in the BAMS detectors, see Figure 7. In case of deviations (either horizontal or vertical) the scanning angles are corrected using a straightforward feedback algorithm as the beam optics simply follows the intercept theorem.

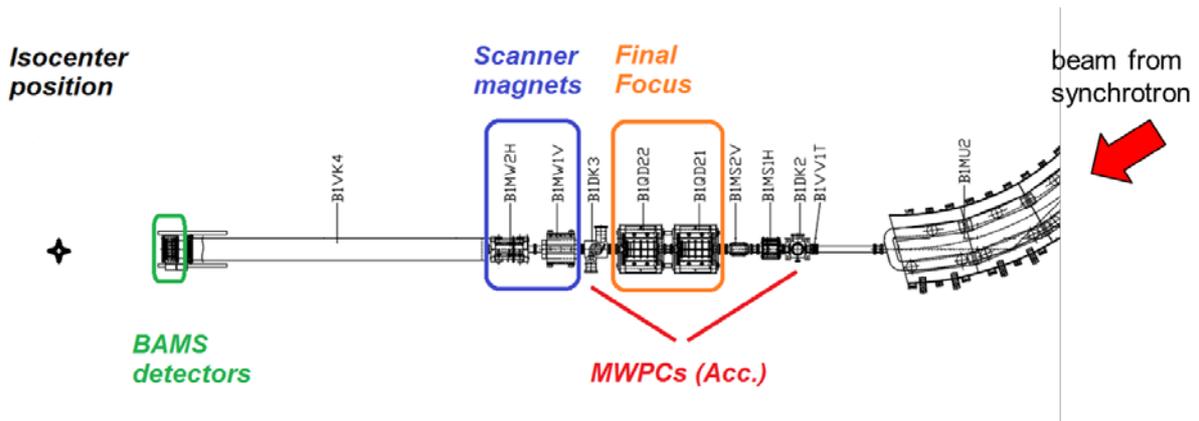

**Fig. 7:** Scheme of a horizontal treatment beam line at HIT showing the built-in elements.

## 3   Dynamic Intensity Control – a Variable Feedback System for Fast Dose Application

As already shown in chapter 2.1 feedback algorithms are implemented in the raster scanning system and thus, the patient therapy works fine, but there are reasons for the reduction of the individual treatment time:

- Higher patient comfort (during treatment the patient is locally immobilized!);
- Higher dose conformity;
- More patients can be treated in the same facility;
- The facility can be operated more economically.

The typical spill structure achieved with the RF knock-out method is shown in Figure 8. An ideal intensity should be completely flat and as close to the limit as possible. In reality the extracted current from the synchrotron shows a different behaviour, the signal has an imperfect macro-structure with a noisy micro-structure. Thus, the scanning velocity is lower than desired and can be enhanced, because the spill quality is essential for the treatment time.

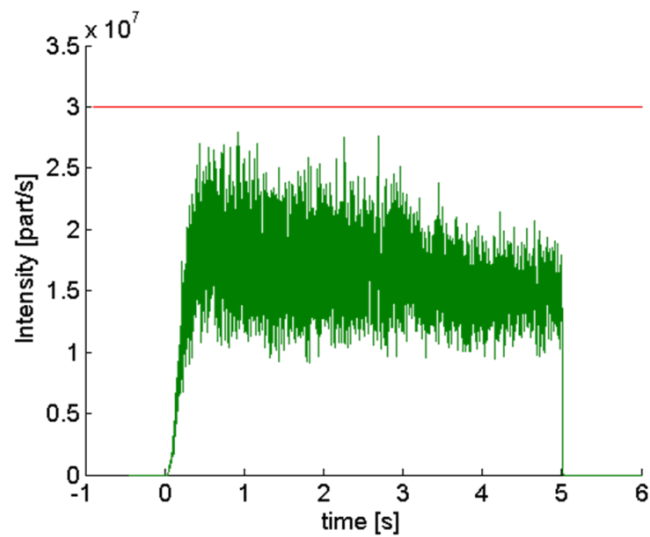

**Fig. 8:** Extracted beam signal from the HIT synchrotron measured by an IC (without feedback).

The idea at this point was to use the intensity signal and add a feedback loop; the implementation is presented in Figure 9 [4].

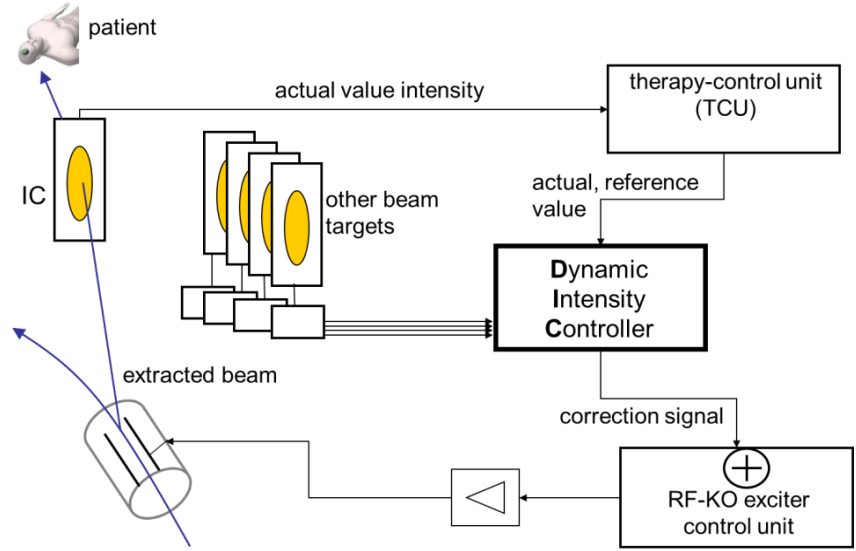

**Fig. 9:** Implementation of the Dynamic Intensity Controller (DIC) as a link between the therapy control system and the RF-KO exciter of the accelerator.

The feedback loop is realized using a PID controller ("P" for a fast response, "I" for no remaining control deviation and "D" optional and currently deactivated). The PID-parameters are a function of the beam parameters ion, energy and intensity. About 10,000 combinations have to be defined, 1% was determined during the commissioning of the DIC; all other values were calculated by spline interpolation. In addition, safe connections have to be installed from the 4 treatment rooms plus beam-dump with their IC detectors to the RF-KO exciter. Additional features could be implemented: a) mitigation of intensity overshoots and b) an "Early abort" function, where the controller realizes when synchrotron is running empty and can signalize this to the therapy control system. Besides the technical issues there was a further challenge: the coupling of the medical product "irradiation system" with the industrial product "accelerator", see also Figure 3.

In a first development step the feedback loop was used to form a more perfect spill structure – the macro-structure looks now more than a rectangle with steep rise in the beginning and nearer to the limit, the micro-structure is less noisy with fewer spikes.

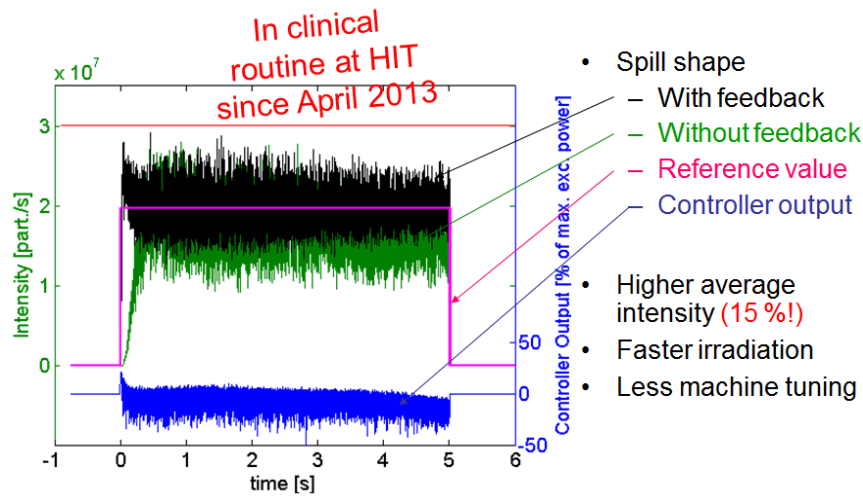

**Fig. 10:** Effects of the active feedback loop (DIC, 1$^{st}$ development step) on the beam quality.

The velocity of the treatment could be increased and the number of interlocks (caused by intensity peaks and other effects) during irradiation was decreased.

But the irradiation time could be furthermore reduced by adaption of the intensity to the actual dose value of each voxel (volume element). The upper diagram in Figure 11 shows an example of a typical dose distribution of one slice. The lowest particle fluency normally determines the intensity for the whole IES. With a fixed intensity the irradiation time per raster point can vary by a factor of > 100, as already mentioned in chapter 2.1. In the 2$^{nd}$ step of the DIC development intensity-modulated spills, where each raster point gets an individual reference value, was implemented.

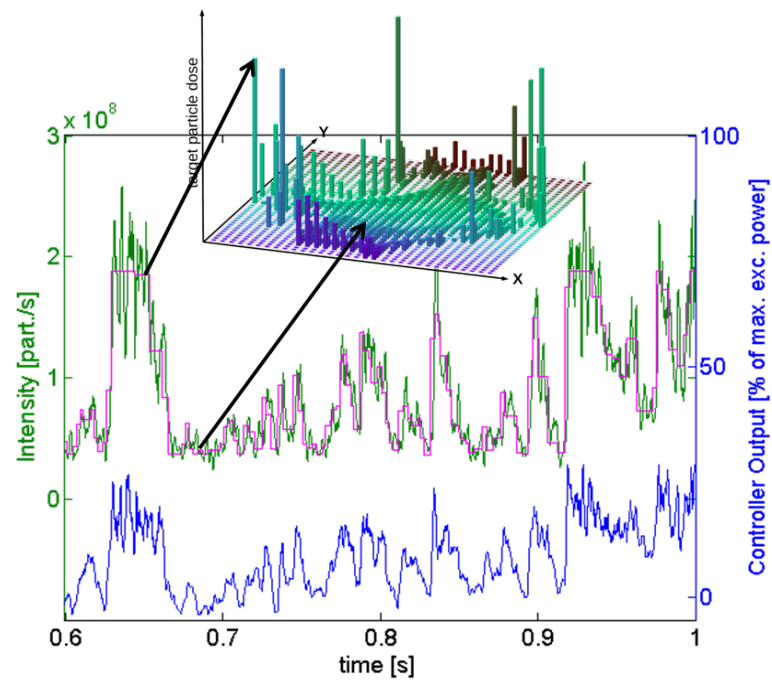

**Fig. 11:** Result of the second DIC development step, see text.

Clearly also this development has its challenges and limitations, as the feedback loop is restricted due to dead times, coming from the following effects:

- Signal detection and digitization, ionization chamber     ~ 150µs
- Particle excitation                                        ~ 0 - 600µs
- Latencies in digital transmission                          ~ 100µs

In addition, irradiating too fast leads to interlocks in the therapy control system and must be avoided. Nevertheless, the feedback loop adapts the actual intensity and thus the beam-on time could again be reduced by ≈45%! This version of the DIC is in clinical routine operation at HIT since April 2014.

## 4    Using Scintillating Screens for Beam Quality Assurance

Viewing Screens are a well-known technique in beam instrumentation since the beginning; the simplest versions are put together from a phosphor screen and a TV camera, newer systems may use more sophisticated scintillating materials and state-of-the-art digital cameras.

Not long ago the medical physicist irradiated films (see Figure 12) and evaluated them for the daily QA to check focus width, 2D „roundness" of the beam as well as homogeneity of the dose application. This takes a lot of time, especially in the case of a gantry, where this had to be done for a lot of angles of incidence with manual adjustment of the films.

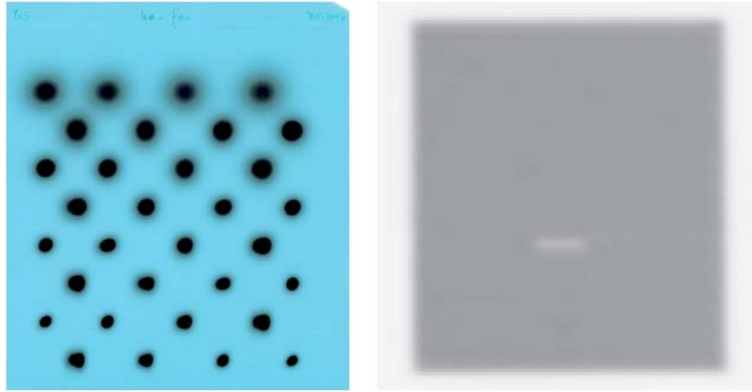

**Fig. 12:** Films irradiated for the daily QA; left film: checking of focus width and "roundness" for different parameters; right film: measurement of the dose homogeneity.

To automatize these procedures ten years ago two different devices were built: a) an adapter robot with rotating mount for films, MWPCs or any other 2D QA measurement devices (→ Patient coordinate system); b) a beam diagnostics for commissioning the beam from the accelerator side – a large scintillating screen fixed to the gantry nozzle (→ Gantry coordinate system), see Figure 13.

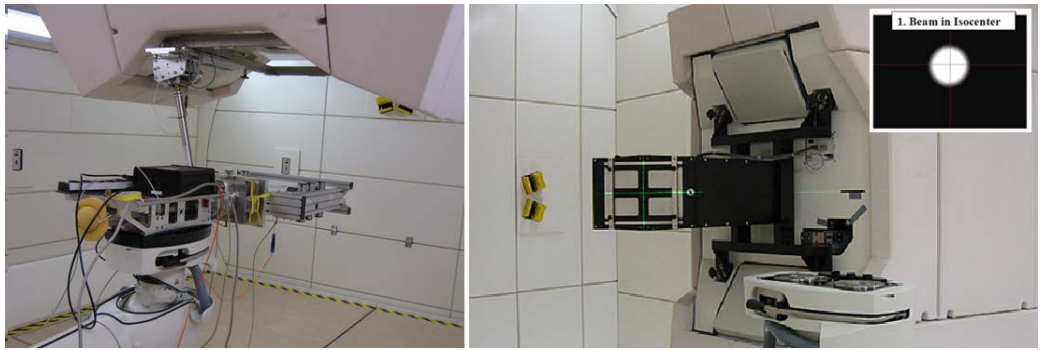

**Fig. 13:** Left photo shows the adapter robot with rotating mount; the right photo shows the large scintillating screen fixed to the gantry nozzle.

But to mount and to adjust these measurement devices is still a lot of work, such a much simpler device was looked for – and found. A system by the company Logos System Int´l [5] for photon and proton beams served as a template for a HIT development – the original system was not specified for all ion beams (He, C, O) with their large dynamic range of intensities. The system is quite simple: It consists of a very thin, light and stiff cone made of carbon fiber material, which contains a thin foil coated with a scintillating phosphor (Lanex® or P43 foils from Proxivision [6]). As in any modern viewing screen system a high-sensitive, cooled CCD or sCMOS MPixel camera looks to this screens and measures the penetrating beam, which gives two beam spots, see Figure 14.

A geometrical reconstruction of the images leads to beam profile data including focal point and focus width. In addition, much more interesting data can be retrieved. For example, evaluating beams passing through the cone from different angles leads to a so-called star plot, which gives the 3D iso-center position.

At HIT the first prototype is still under development [7]; all HW parts of the measurement device are defined meanwhile, but now the triggered image acquisition and the evaluation program have to be coded, such that the system can be used as a medical product.

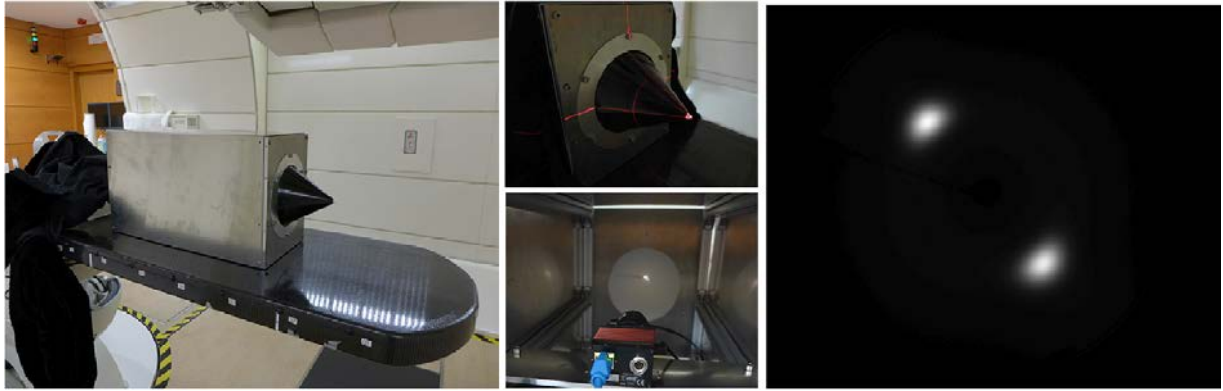

**Fig. 14:** Prototype device at HIT gantry with cone for the beam passage (left side); Adjustment of the cone with a 3D laser system (middle top); Camera and cone from inside (middle bottom); Photo of penetrating proton beam with 106.8 MeV and a focus width of 15 mm (right side).

## 5 Combining Particle Therapy and MRI

A strong motivation for doctors is: "Seeing what you treat", that means an online diagnostics of the irradiated tumor would be favorable [8]. A CT in the treatment room is almost standard today, but a MRI would cause no further radiation dose, which is especially important in pediatric treatments. Furthermore, the observation of possible tumor shrinkage during therapy is necessary to avoid errors in the adapted dose allocation, see Figure 15 for an example.

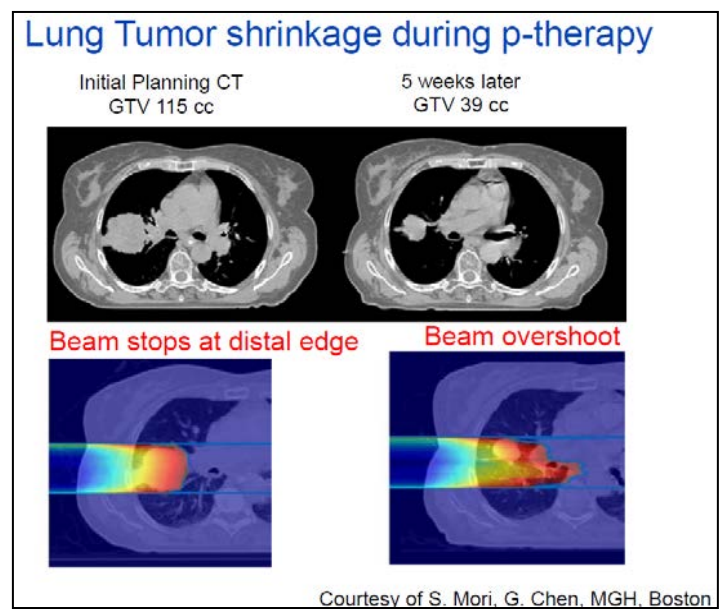

**Fig. 15:** An example of a tumour shrinkage leading to beam overshoot.

Additionally, doctors normally use large margins around the tumor volume because of possible range uncertainties, movement of the tumor along the irradiation, etc. – but in this way more healthy tissue will be irradiated than necessary. For example: A 6.5 mm thick margin consists of the same volume as a 5 cm diameter target. Thus, better diagnostics could lead to shrinkage of these margins.

MRI-Linac systems (with photon beams) are currently being introduced in radiotherapy, a cross section of such a combined MRI and photon linac (MRIdian® Linac with 0.35 T magnetic field) is shown in Figure 16 (left picture). MRI devices have very complex magnetic fields (see Figure Fig. 16, right sketches), which – in the case of particle therapy – would interact with proton and/or ion beams.

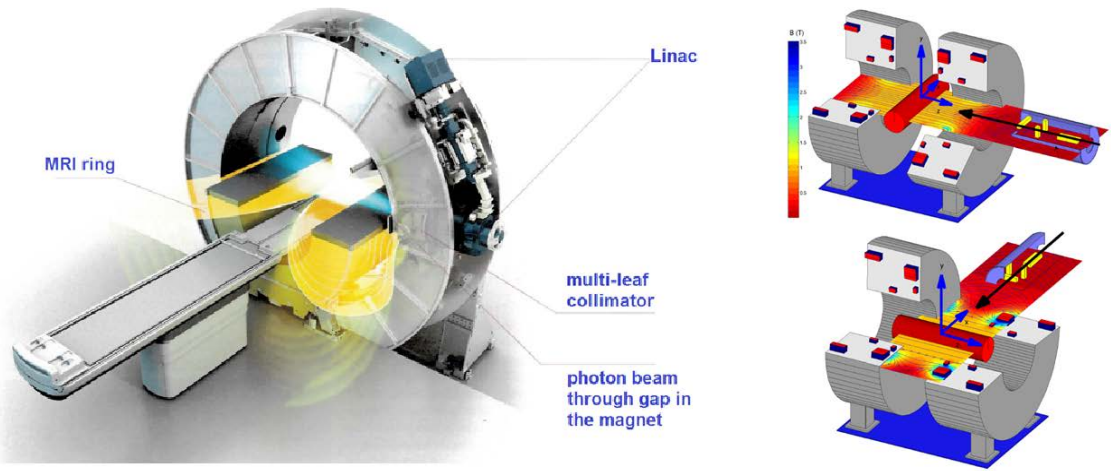

**Fig. 16:** MRI-Linac system (left side); possible arrangements the magnetic field of the MRI and the beam directions (right side)

In case of a diverging point source (DPS), e.g. a raster scanning system, complex distortions and deflections of the beam occur [9], with several mm up to cm much too large to be accepted, see Figure 17. Furthermore, also effects on charged secondary particles produced during the irradiation have to be taken into account. And the influence of pulsed magnetic fields from the MRI has to be carefully examined! Complex mechanisms like compensation procedures using analytic algorithms or look-up tables will be necessary. Alternatively, one could think of active feedback systems.

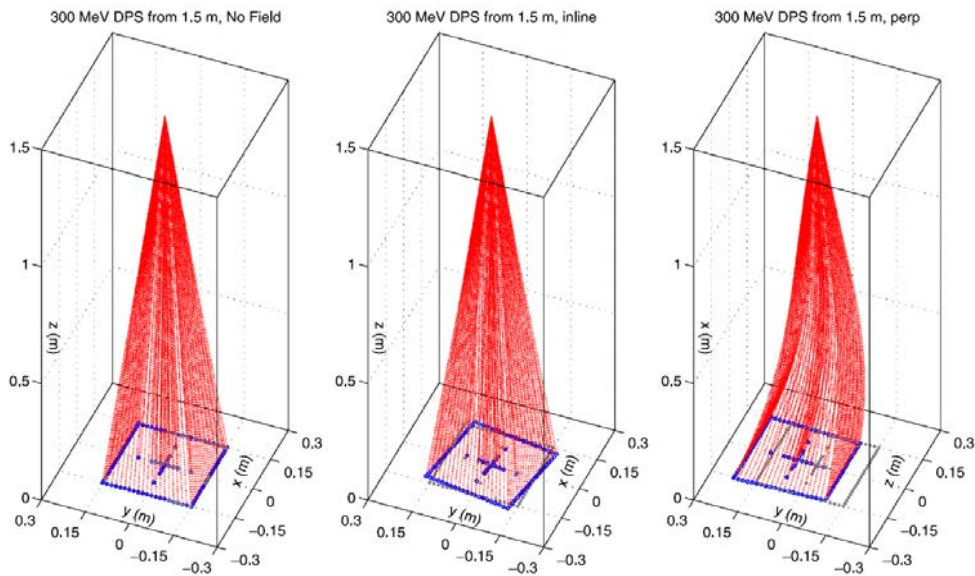

**Fig. 17:** Simulation of a beam in a raster scanning system influenced by a magnetic field by a MRI; left: without field, middle: with inline field, right: with perpendicular field.

In any case it needs beam intensity and profile measurements in magnetic fields, the nowadays used methods using ionization chambers or MWPCs will be strongly influenced, maybe they will not measure correctly anymore. So other diagnostics are under investigation now. One could learn from large detectors for nuclear physics experiments, which use magnetic fields, but have to take into account that the readout schemes are totally different as well as the beam intensities and its dynamics in particle therapy. Possible solutions are still under study.

It is obvious that the development process of a combined system of MRI and particle therapy is rather complex. An iteration process is necessary: it starts with magnetic field and beam dynamics simulations (like in Figure 17), followed by an analysis of the influence on the scanned beam and the beam diagnostics. In the next step a feedback system has to be simulated. In parallel, a disturbance-free BAMS system hast to be developed. At last, an experimental verification has to be set up and all energy and intensity dependencies have to be measured. It may be necessary to go several times through this process to create a system which can be medically approved and certified at the end before used for patient treatment. Beam instrumentation is an important and integral part of this process, which is still under development.

## 6    Conclusion and Acknowledgements

There is no difference between a scientifically used accelerator and a medical accelerator concerning beam diagnostics equipment, only economic effects may play a role concerning the number and variety of the instrumentation. The specialty of medical accelerators is the strong link to the *beam consumer*, in this case the irradiation technology and the high-conformal dose application. The common topics "Instrumentation – Feedbacks – Detector Technology – Quality Assurance" lead to close collaboration of accelerator and medical physics people in such a facility, resulting in new and innovative tools for dose application, beam tuning and daily QA as shown in this talk.

Many thanks go to all colleagues from HIT – especially to Thomas Haberer, Jakob Naumann, Christian Schömers, Oliver Jäckel, Julian Horn and Jochen Schreiner – for their advice and providing me interesting slides, example measurement results, photos and literature links.

## 7    References

[1] A. Peters, P. Forck, Beam Diagnostics for the Heavy Ion Cancer Therapy Facility (HICAT), Beam Instrumentation Workshop, May 2000, Boston, MA, USA, AIP Conference Proceedings 546. pp. 519-526.

[2] Th. Haberer, W. Becher, D. Schardt, G. Kraft, Magnetic scanning system for heavy ion therapy, Nucl. Instr. Meth. A330, pp. 296 – 305 (1993).

[3] Th. Haberer, J. Debus, H. Eickhoff, O. Jäkel, D. Schulz-Ertner, U. Weber, The Heidelberg Ion Therapy Center, Rad. Onc. 73 (2), pp. 186 – 190 (2004).

[4] C. Schoemers et al., The intensity feedback system at Heidelberg Ion-Beam Therapy Centre, Nucl. Instrum. Methods Phys. Res., Sect. A 795 (2015), pp. 92–99.

[5] www.logosvisionsystem.com

[6] www.proxivision.de

[7] Harald Latzel et al., Usage of scintillation screens at the medical facility HIT, Joint ARIES-ADA Topical Workshop on Scintillation Screens and Optical Technology for transverse Profile Measurements, April 2019, see https://indico.cern.ch/event/765975/contributions/3370885/

[8] B. Oborn et al., MRI Guided Proton Therapy: Pencil beam scanning in an MRI fringe field, International Conference on Translational Research in Radio-Oncology, Geneva, 2016, see https://indico.cern.ch/event/392209/contributions/1828309/attachments/1246384/1835778/pbs_in_mrpt.pdf

[9] B. Oborn et al., Medical Physics, 42, 2113 (2015).